\documentclass[a4paper,fleqn,usenatbib]{mnras}
\usepackage{newtxtext,newtxmath}
\usepackage[T1]{fontenc}
\usepackage{ae,aecompl}
\usepackage{graphicx}	
\usepackage{amsmath}	

\usepackage{amssymb}	
\usepackage{mathrsfs}
\usepackage{longtable}
\usepackage{hyperref}
\usepackage[font=small]{caption}
\usepackage{bm}

\title[A late-time source of AT2018cow]{A hot and luminous source at the site of the fast transient AT2018cow at 2--3 years after its explosion}

\author[N.-C. Sun et al.]{Ning-Chen Sun$^1$\thanks{E-mail: n.sun@sheffield.ac.uk}, Justyn R. Maund$^1$, Paul A. Crowther$^1$ and Liang-Duan Liu$^2$ \\
1 Department of Physics and Astronomy, University of Sheffield, Hicks Building, Hounsfield Road, Sheffield S3 7RH, UK \\
2 College of Physical Science and Technology, Central China Normal University, 152 Luoyu Road, Wuhan, Hubei 43079, China}

\date{Accepted XXX. Received YYY; in original form ZZZ}

\pubyear{2022}

\begin{document}
\label{firstpage}
\pagerange{\pageref{firstpage}--\pageref{lastpage}}
\maketitle

\defcitealias{ck04.ref}{CK04}

\begin{abstract}

We report the discovery of a luminous late-time source at the position of the fast blue optical transient (FBOT) AT2018cow on images taken by the Hubble Space Telescope (HST) at 714~d and 1136~d after its explosion. This source is detected at both UV and optical wavelengths and has prominent H$\alpha$ emission. It has a very stable brightness between the two epochs and a very blue spectral energy distribution (SED) consistent with $f_\lambda$~$\propto$ $\lambda^{-4.1 \pm 0.1}$, i.e. the Rayleigh-Jeans tail of a hot blackbody with a very high temperature of log($T$/K)~$>$ 4.6 and luminosity of log($L$/$L_\odot$)~$>$ 7.0. This late-time source is unlikely to be an unrelated object in chance alignment, or due to a light echo of AT2018cow. Other possible scenarios also have some difficulties in explaining this late-time source, including companion star(s), star cluster, the survived progenitor star, interaction with circumstellar medium (CSM), magnetar, or delayed accretion in a tidal disruption event (TDE). Long-term and multi-wavelength monitoring will help to resolve its nature and finally reveal the origin of the ``Cow".

\end{abstract}


\defcitealias{P19}{P19}
\defcitealias{X21}{X21}

\section{Introduction}
\label{intro.sec}

FBOTs have peak brightnesses comparable to those of (typical or superluminous) supernovae (SNe). As implied by their name, however, they exhibit very blue colors and evolve with much shorter timescales \citep{Ho2021}. Their rapid evolution is hard to explain with radioactive decay that powers most SNe, suggesting that they are a distinct new type of transients (e.g. \citealt{X21}, hereafter \citetalias{X21}). At a distance of only 63~Mpc, AT2018cow is the closest FBOT ever discovered. It has a peak luminosity of $\sim$10$^{44}$~erg~s$^{-1}$, rises to peak in just a few days and then declines dramatically (\citealt{P18}; \citealt{P19}, hereafter \citetalias{P19}; \citealt{X21}). The photospheric temperature is $\sim$30,000~K near peak and still as high as $\sim$10,000~K at $\sim$50~d after explosion (\citealt{Kuin2019}; \citetalias{P19, X21}).

It is still unclear what progenitor is responsible for AT2018cow and what process(es) powers its rapid evolution. Current models include TDE (\citealt{Liu2018, Kuin2019}; \citetalias{P19}), CSM interaction (\citealt{Fox2019}; \citetalias{X21}), magnetar (\citealt{P18}; \citealt{Fang2019}), pulsational pair-instability SN \citep[PPISN;][]{Leung2020}, jet-envelope interaction in a core-collapse SN \citep{jet.ref}, or a jet-driven SN impostor from a common-envelope binary system \citep{Soker2022}. It is worth mentioning that AT2018cow is located in an environment typical for core-collapse SNe \citep{L20}; more recently, \citet{Pasham2021} found evidence for a rapidly spinning compact object, which could be a neutron star or a black hole of mass $<$850~$M_\odot$.

For a number of SNe, late-time observations sometimes reveal significant brightness unexpected from their early-time evolution. The late-time sources contain important information for these SNe as their possible light echos \citep[e.g. SN~2011dh;][]{Maund2019}, binary companions \citep[e.g. SN~1993J and SN~2006jc;][]{Maund2004, Maund2016, Sun2020a}, host star clusters \citep[e.g. SN~2014C;][]{Sun2020b}, late CSM interaction \citep[e.g. SN~1993J;][]{Zhang2004} or something else. In comparison, no late-time sources have ever been reported for FBOTs, most of which are located in distant galaxies and observations become very difficult beyond several months after their explosions.

Thanks to its relative proximity, AT2018cow may be the only possible FBOT to study its brightness at significantly late times. In this paper, we report the discovery of a bright source at the position of AT2018cow at 2--3~years after its explosion. This is somewhat surprising for a fast evolving transient with a steeply declining light curve. We describe the observed features and then discuss its possible origin scenarios.

Throughout this paper, we use a redshift of 0.01406 and a distance of 63~Mpc for the host galaxy \citepalias[CGCG~137-068;][]{X21} and a Galactic reddening of $E(B-V)$~= 0.078~mag for AT2018cow \citep{galebv.ref}. Its internal reddening within the host galaxy is negligible \citepalias{P19, X21}. All epochs are relative to an estimated explosion date of MJD~= 58,284.79 \citepalias{X21}.

\section{Data}
\label{data.sec}

\begin{table}
\caption{HST/WFC3/UVIS observations of AT2018cow.}
\begin{tabular}{ccccc}
\hline
\hline
Program & Epoch$^{\rm d}$ & Filter & Exposure & Magnitude$^{\rm e}$ \\
ID & (day) & & Time (s) & (mag) \\
\hline
15600$^{\rm a}$
& 52 & F218W &  880 & 19.12 (0.02) \\
& 52 & F225W &  770 & 18.99 (0.01) \\
& 52 & F275W &  280 & 18.88 (0.02) \\
& 52 & F336W &  150 & 19.17 (0.02) \\
\hline
15600
& 57 & F218W &  880 & 19.58 (0.01) \\
& 57 & F225W &  770 & 19.48 (0.01) \\
& 57 & F275W &  280 & 19.34 (0.01) \\
& 57 & F336W &  150 & 19.58 (0.01) \\
\hline
15600
& 62 & F218W &  880 & 19.75 (0.03) \\
& 62 & F225W &  770 & 19.56 (0.01) \\
& 62 & F275W &  280 & 19.56 (0.02) \\
& 62 & F336W &  150 & 19.82 (0.02) \\
\hline
15974$^{\rm b}$
& 714 & F657N &  1119 & $>$24.65 (5$\sigma$) \\
& 714 & F665N &  1119 & 24.47 (0.24) \\
& 714 & F225W &  1116 & 22.55 (0.06) \\
& 714 & F336W &  1116 & 23.32 (0.05) \\
& 714 & F555W &  1044 & 25.64 (0.07) \\
& 714 & F814W &  1044 & 25.82 (0.19) \\
\hline
16179$^{\rm c}$
& 1136 & F555W &  710 & 25.63 (0.08) \\
& 1136 & F814W &  780 & 25.96 (0.24) \\
\hline
\end{tabular} \\
PIs: (a) Foley R.; (b) Levan A.; (c) Filippenko A. \\
(d) Epoch is relative to an explosion date of MJD~= 58,284.79 \citepalias{X21}. \\
(e) All magnitudes are in the Vega system.
\label{obs.tab}
\end{table}

AT2018cow was observed by three HST programs (Table~\ref{obs.tab}), all conducted with the Ultraviolet-Visible (UVIS) channel of the Wide Field Camera 3 (WFC3). The first program (ID: GO-15600) was performed at $t$~= 52, 57 and 62~d while the other two programs (IDs: 15974 and 16179) were carried out at significantly later times of $t$~= 714 and 1136~d, respectively. We retrieved the images from the Mikulski Archive for Space Telescopes (\url{https://archive.stsci.edu/index.html}) and re-drizzled them with \texttt{driz\_cr\_grow = 3} for better cosmic ray removal (all other parameters were left unchanged as in the standard calibration pipeline).

In this work, we also use AT2018cow's early-time light curves of \citetalias{P19} out to $t$~$\gtrsim$ 60~days. At optical wavelengths, the reported magnitudes in the Swift/UVOT and SDSS-like filters were converted into the Johnson-Cousins $UBVRI$ system, using the relative offsets shown in Fig.~2 of \citetalias{P19}. Magnitudes in the $V$ and $I$ bands are very similar to those in the F555W and F814W bands, respectively, with differences no larger than the photometric uncertainties \citep{Harris2018}. Light curves in the Swift/UVOT UVM2 and UVW1 filters and in the $U$ band were transformed to the HST F218W, F225W, F275W and F336W bands using relations derived with synthetic magnitudes of blackbody spectra of 7000--50,000~K (calculated with the \textsc{pysynphot} package). \citetalias{P19} used the AB magnitude system and we converted their magnitudes into the Vega magnitude system with $m_{\rm AB} - m_{\rm Vega}$ = 1.68, 1.66, 1.50, 1.19, $-$0.02 and 0.43~mag for the F218W, F225W, F275W, F336W, F555W and F814W bands, respectively. These offsets were also calculated with the \textsc{pysynphot} package and are valid for hot sources like AT2018cow.

\section{Detection of a late-time source}
\label{source.sec}

\begin{figure*}
\centering
\includegraphics[width=1\linewidth, angle=0]{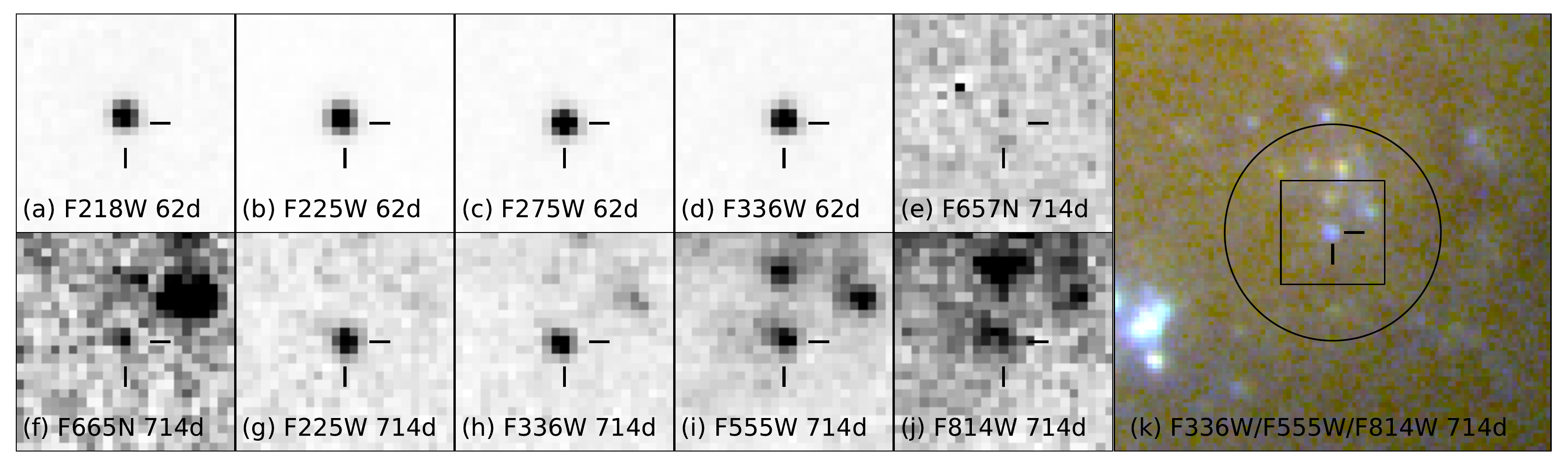}
\caption{Example images of AT2018cow at early and late times of $t$~= 62~d (a--d) and 714~d (e--j), respectively; each panel has a dimension of 1"~$\times$ 1" and is centered on AT2018cow (shown by the cross hair). No source is detected at the position of AT2018cow in the F657N image (e). In the F814W image (j), the late-time source is not very obvious by eye due to a bright neighboring star to its northeast. (k) Three-color composite of the F336W, F555W and F814W images at $t$~= 714~d; the square corresponds to the extent of the other panels (a--j) and the circle has a radius of 1". All panels are aligned with North up and East to the left. An angular size of 1" corresponds to a linear size of 305~pc at the distance of AT2018cow.}
\label{image.fig}
\end{figure*}

We performed point-spread-function (PSF) photometry on the HST images with the \textsc{dolphot} package \citep{dolphot.ref}. At the position of AT2018cow, a point source is significantly detected in all the broad-band images, only marginally detected in the F665N image with a 4.6$\sigma$ significance, and not detected in the F657N image (Fig.~\ref{image.fig}). The magnitudes or magnitude limits (obtained with artificial star tests) for this source are listed in the last column of Table~\ref{obs.tab}. Note that the  late-time source is not spatially resolved; therefore, its size should not significantly exceed that of the PSF, which corresponds to $\sim$20~pc at the distance of AT2018cow.

\paragraph*{Light curves}

\begin{figure}
\centering
\includegraphics[width=0.95\linewidth, angle=0]{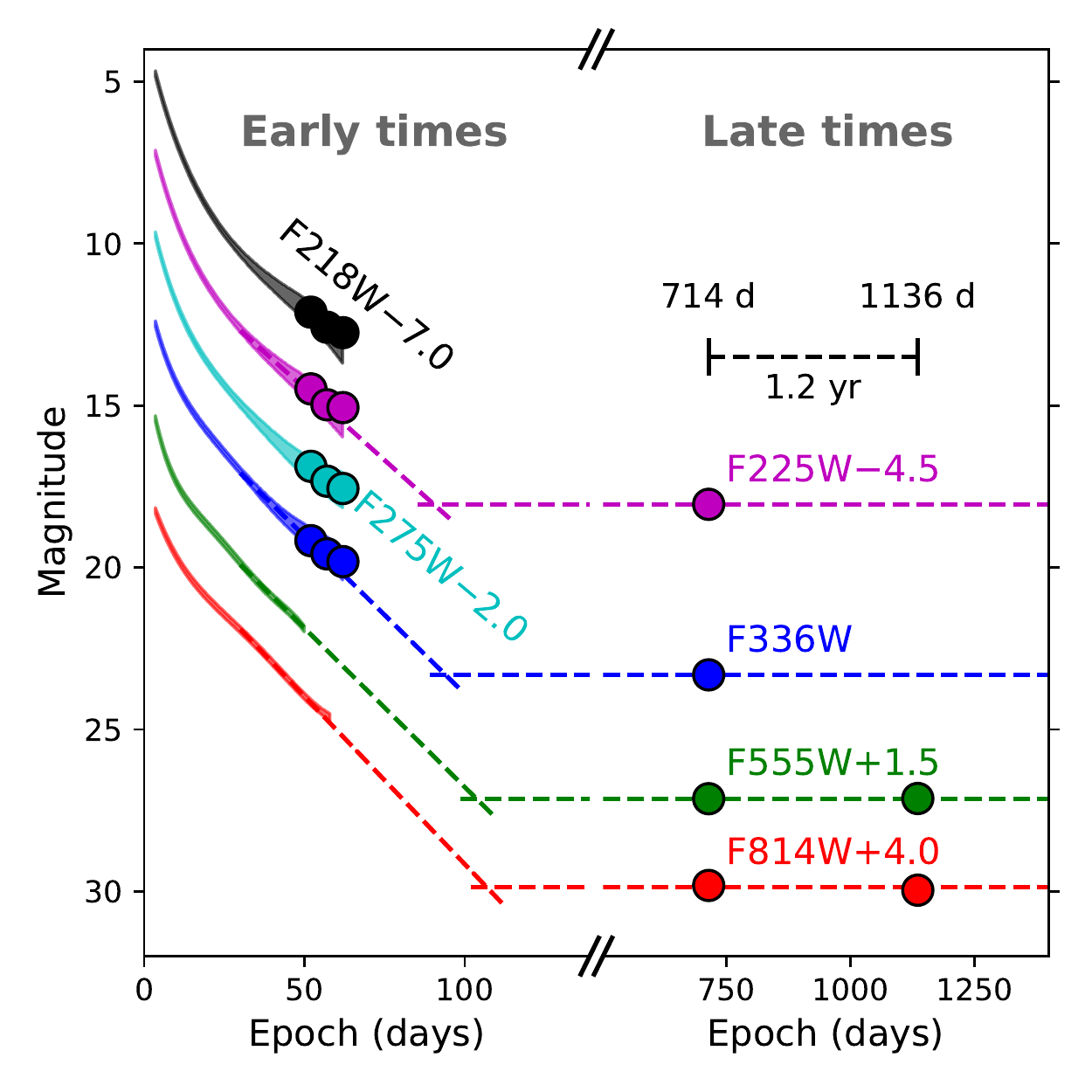}
\caption{HST photometry of AT2018cow (filled circles) with error bars smaller than the symbol size. The solid lines are AT~2018cow's early-time light curves reported by \citetalias{P19} (with line thickness showing the photometric uncertainties), which have been converted into the HST filters and the Vega magnitude system. The dashed lines correspond to linear extrapolations of the light curve tails or the brightness of the late-time source.}
\label{curve.fig}
\end{figure}

At the first three epochs of the HST observations ($t$~= 52, 57 and 62~d), the derived magnitudes are consistent with the light curves of \citetalias{P19} within the uncertainties (Fig.~\ref{curve.fig}). AT2018cow's brightness evolves very rapidly; if we simply extrapolate the light curve tails, the brightness should decline to a very low level within a few months. Somewhat surprisingly, there is still a bright source at the position of AT2018cow at $t$~= 714 and 1136~d, or $\sim$2--3 years after its explosion. It is also worth noting that the magnitudes of the late-time source are strikingly stable, in both F555W and F814W bands, over a time span of $\sim$1.2~yr between the last two epochs. The magnitude differences are only 0.01~mag and 0.14~mag, respectively, much smaller than the photometric uncertainties.


\paragraph*{SED}

\begin{figure}
\centering
\includegraphics[width=1\linewidth, angle=0]{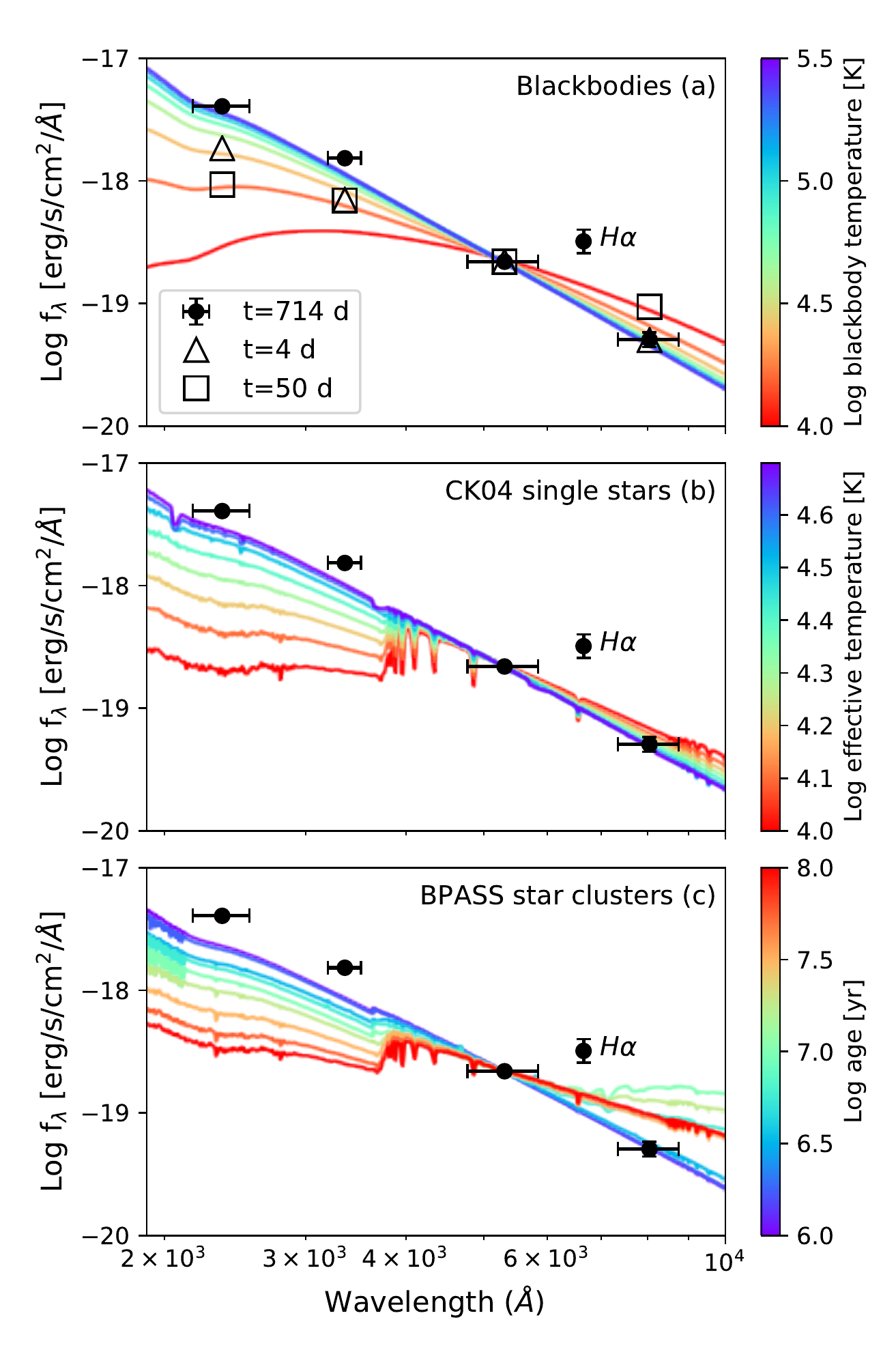}
\caption{SED of AT2018cow's late-time source (black data points); the horizontal error bars correspond to the root-mean-square widths of the HST filters and the vertical error bars reflect photometric uncertainties (if not smaller than the symbol size). The F657N detection limit does not provide additional constraint and is omitted in the plot. For comparison we show model spectra (colored lines) for (a) blackbodies, (b) single stars \citep{ck04.ref}, and (c) star clusters \citep[from \textsc{bpass} v2.1 binary population synthesis;][]{bpass.ref}, all reddened with AT2018cow's Galactic reddening and normalized to the F555W band. In (a) we also show the early-time SED of AT2018cow at $t$~= 4 d (open triangles) and 50 d (open squares) \textbf{normalized to the F555W band}; their error bars are no larger than the symbol size.}
\label{sed.fig}
\end{figure}

The late-time source has an SED even bluer than AT2018cow at early times (\citealt{P18}; \citealt{Kuin2019}; \citetalias{P19, X21}). In Fig.~\ref{sed.fig} the observed SED is compared with model spectra for blackbodies, single normal stars \citep{ck04.ref} and star clusters with a \citet{imf.ref} initial mass function (IMF) from \textsc{bpass} v2.1 binary population synthesis models \citep{bpass.ref}. The late-time source has significant UV excess even compared with the hottest stars of 50,000~K\footnote{If a star has strong wind, its continuum would appear redder and contamination due to emission lines is more significant for the optical filters than for the UV filters \citep{Gotberg2017}. Therefore, this effect is not able to explain the UV excess. We note, however, that the helium-burning Wolf-Rayet stars may reach significantly higher temperatures, up to 10$^5$~K \citep{Crowther2007}, than the hottest stars included in the \citet{ck04.ref} models.} or the youngest star clusters of 1~Myr\footnote{We also checked the effect of stochastic sampling on star clusters' SEDs with the \textsc{slug} package \citep{slug1.ref, slug2.ref}. This effect may make the SEDs even redder, due to some massive stars having evolved into red supergiants, and thus more inconsistent with observations.}. We find a power-law solution of $f_\lambda$~$\propto$~$\lambda^{-4.1 \pm 0.1}$ by fitting to the data, consistent with the Rayleigh-Jeans tail of very hot blackbodies.

\paragraph*{Position on the HR diagram}

\begin{figure}
\centering
\includegraphics[width=1\linewidth, angle=0]{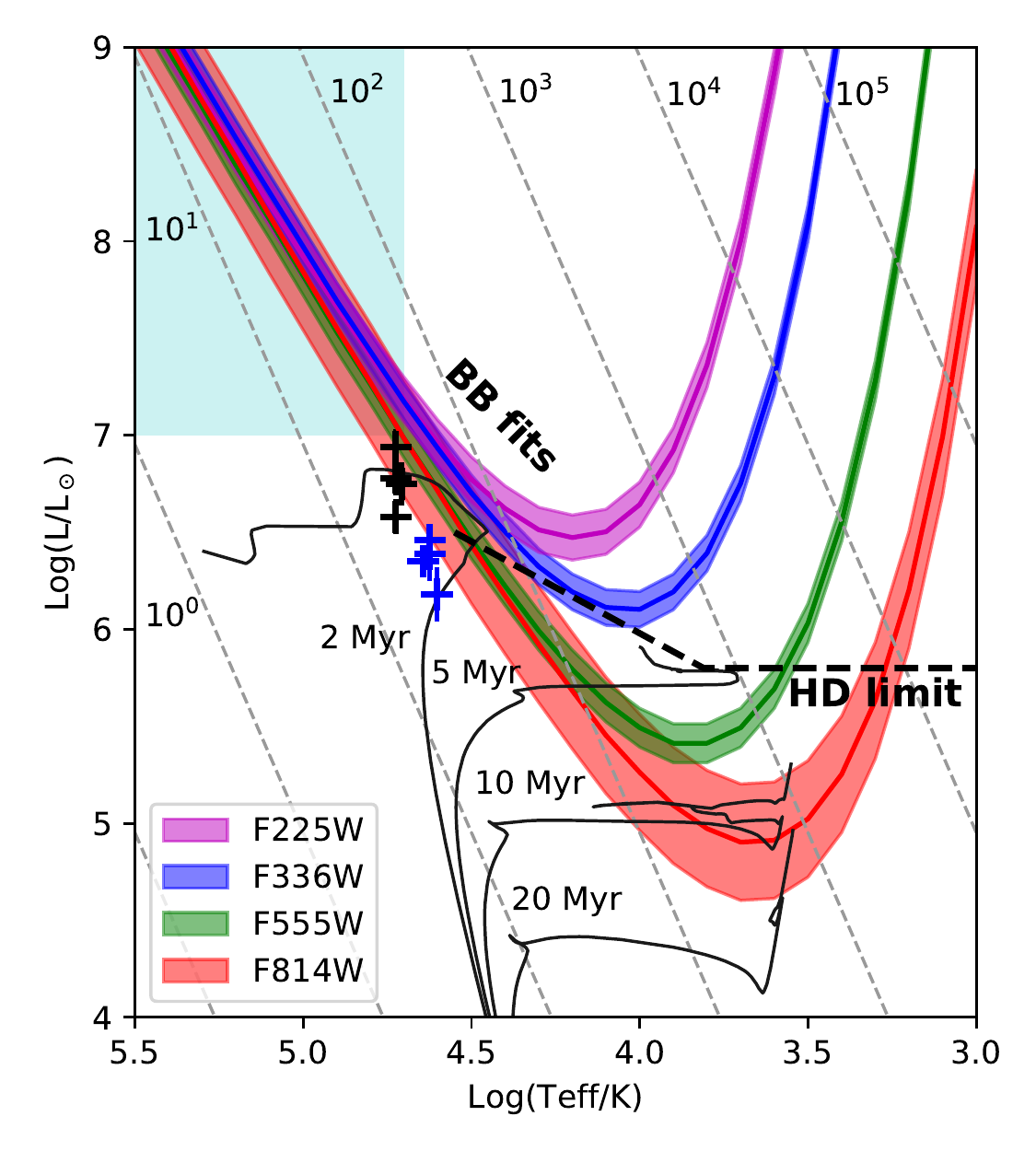}
\caption{In this HR diagram, we use a color-shaded region, for each HST band, to show the possible positions of blackbodies whose synthetic magnitudes match the observed late-time brightness of AT2018cow within 5$\sigma$ uncertainties. The four regions should converge together if the observed SED is indeed a blackbody spectrum across all four bands. This is met at log($T_{\rm eff}$/K)~$>$ 4.7 and log($L$/$L_\odot$)~$>$ 7.0, i.e. the light blue-shaded area in the upper-left corner. The grey dashed lines correspond to constant blackbody radii and the labelled values are in units of solar radius ($R_\odot$). The thin black solid lines are \textsc{parsec} v1.2S single stellar isochrones \citep{parsec.ref} and the thick black dashed line is the \citet{hdlimit.ref} limit for stellar luminosity. The ``$+$" symbols show some of the most luminous stars in the Milky Way (NGC~3603-A1a, A1b, B and C; in blue) and the Large Magellanic Cloud (R136a1--3 and b; in black; \citealt{Crowther2010}).}
\label{hrd.fig}
\end{figure}

In Fig.~\ref{hrd.fig} we show the possible positions of blackbodies on the HR diagram, whose synthetic magnitudes can match the observed ones for the late-time source. The result suggests that, if the late-time source is indeed a blackbody, it should have a very high temperature of log($T_{\rm eff}$/K)~$>$ 4.7 and luminosity of log($L$/$L_\odot$)~$>$ 7.0, corresponding to a blackbody radius of several tens of solar radii. The derived luminosity is significantly higher than any typical single stars, which is obvious by comparing it with the \citet{hdlimit.ref} limit\footnote{Recently, \citet{Davies2018} found evidence for an even lower luminosity limit for cool supergiants in the Magellanic Clouds.}, the \textsc{parsec} single-stellar isochrones \citep{parsec.ref}, and some of the most luminous very massive stars of $>$100~$M_\odot$ in the Milky Way or in the Large Magellanic Cloud (\citealt{Crowther2010}; see also \citealt{Bestenlehner2020}). Note, however, that the bolometric luminosity remains very uncertain since it is not clear whether the late-time source deviates from a blackbody SED outside the observed wavelength range.

\paragraph*{H$\alpha$ emisssion}

The F665N detection indicates prominent H$\alpha$ emission from the late-time source (Fig.~\ref{sed.fig}). Note that the wavelength of H$\alpha$ is redshifted away from F657N into the F665N band due to the recession of the host galaxy. By subtracting a continuum interpolated between the F555W and F814W bands, we estimate a wavelength-integrated line flux of $\sim$ 10$^{-17}$~erg~s$^{-1}$~cm$^{-2}$ or an H$\alpha$ luminosity of  $\sim$4~$\times$ 10$^{36}$~erg~s$^{-1}$ (corrected for extinction). It is not clear whether the H$\alpha$ emission arise from stellar sources (e.g. WNh stars; \citealt{Schaerer1998}) and/or from a modest and compact H~\textsc{ii} region at the position of AT2018cow (for comparison, H~\textsc{ii} regions have H$\alpha$ luminosities ranging from 10$^{37}$~erg~s$^{-1}$ for the Orion Nebula to $>$10$^{40}$~erg~s$^{-1}$ for 30~Doradus; \citealt{Crowther2013}). With integral-field spectroscopy, \citet{L20} found AT2018cow to be associated with a giant H~\textsc{ii} region. The F665N image, which has a much higher spatial resolution, shows that the giant H~\textsc{ii} region is actually to the northwest of AT2018cow with an offset of $\sim$100~pc. Therefore, the detected F665N source is not due to the giant H~\textsc{ii} region in the environment of AT2018cow.

\section{Possible origins}
\label{origin.sec}

In this section we discuss what objects or physical mechanism(s) may give rise to the late-time brightness of AT2018cow.

\paragraph*{Chance alignment}
The detected late-time source is less likely due to an unrelated object in chance alignment. We performed relative astrometry with 11 reference stars between the late-time images and the F218W image at $t$~= 62~d. The (transformed) positions have an offset of only 0.3~pixel between AT2018cow and the late-time source, smaller than the astrometric uncertainty of 0.5~pixel. Additionally, there are only 22 sources significantly detected within 1" from AT2018cow (Fig.~\ref{image.fig}k); the probability is $<$1\% for a randomly positioned source to coincide with AT2018cow within the 1$\sigma$ error radius. Moreover, very few sources in the surrounding region can match AT2018cow's late-time brightness and its very blue SED. Therefore, we suggest the late-time source is not in chance alignment but physically associated with AT2018cow.

\paragraph*{Light echo}
It is very difficult to explain AT2018cow's late-time brightness with a light echo. The H$\alpha$ emission is inconsistent with AT2018cow's featureless spectrum near peak \citepalias{P19, X21}, and it may require a very specific dust configuration to reproduce its very blue SED. For circumstellar dust from a steady wind, the brightness of its light echo declines continuously  \citep{Chevalier1986}, inconsistent with the observed stable F555W/F814W brightness from $t$~= 714~d to 1136~d. A foreground dust sheet may produce a stable light echo brightness if it is distant enough from the transient \citep{Cappellaro2001}. However, AT2018cow has a negligible amount of dust in its foreground since observations find very little extinction from within the host galaxy (\citetalias{P19, X21}). For dust in the background, the scattering is much less efficient due to the large scattering angles \citep{Draine2003}. Therefore, it seems unlikely that the late-time source is a light echo.

\paragraph*{Stellar source(s)}

A late-time source with stable brightness is naturally expected if the transient's progenitor is located inside a binary/multiple system \citep{Maund2004, Maund2016, Sun2020a, Sun2022} or star cluster \citep{Sun2020b}. For AT2018cow, the late-time source is inconsistent with a single companion star; the very high luminosity requires it to be at least 2--3 very massive ($>$100~$M_\odot$) stars inside a multiple system (Fig.~\ref{hrd.fig}). Star clusters with standard IMFs are difficult to match the observed SED (Fig.~\ref{sed.fig}); if the late-time source were indeed a star cluster, it may have a very top-heavy IMF \citep[e.g.][]{30dor.ref} so that the brightness is dominated by the very massive stars.

It is unclear if FBOTs mark the death of massive stars or arise from their non-terminal explosions. In a PPISN, for example, material ejected at a later epoch may collide with earlier ejecta, producing a luminous transient, but the progenitor may not have completely destroyed itself and can still be observed at late times \citep{Woosley2007}. It is, however, still difficult to explain the high luminosity of AT2018cow's late-time source unless the progenitor is overluminous due to some special mechanisms. In a PPISN model of AT2018cow, \citet{Leung2020} found a progenitor mass of $M_{\rm ini}$~$\sim$ 80~$M_\odot$ in order to fit the fast evolving light curves.

In these scenarios, the H$\alpha$ emission may arise from stars with strong winds or from the local ionized gas, and AT2018cow is (probably) related to a very young and massive progenitor. Caution, however, that some SNe spatially aligned with young star clusters are found to have much older progenitors due to sequentially triggered star formation \citep[e.g. SN~2012P;][]{Sun2021}.

\paragraph*{Late emission from AT2018cow}

AT2018cow has been proposed to be a core-collapse SN powered by CSM interaction (\citealt{Fox2019}, \citetalias{X21}). In the model of \citetalias{X21}, AT2018cow was surrounded by CSM with a flat density profile from 3 to 1200~$R_\odot$. If the late-time source were also due to CSM interaction, then the CSM should extend out to a much larger distance of $\gtrsim$~5~$\times$ 10$^{16}$~cm (for an ejecta velocity of 5000~km/s); the CSM density may also be different at larger distances in order to explain the significant light curve flattening from early to late times. In this scenario, the late-time emission may be contributed by emission lines. However, it may be challenging to reproduce the very stable brightness from $t$~= 714 to 1136~d; most interacting SNe have time-varying brightness and their light curve plateaus, if any, last for much shorter periods (\citealt{Mauerhan2013a, Mauerhan2013b}; but see \citealt{Smith2009} for the exceptions of extremely enduring SN~1988Z and SN~2005ip).

In a magnetar-powered model of AT2018cow, \citet{P18} derived a spin period of 11~ms and a magnetic field strength of 2.0~$\times$ 10$^{15}$~G, corresponding to a magnetic dipole spin-down timescale of 0.3~d \citep{Fang2019}. In this case, the energy injection rate is expected to decline by 60\% from $t$~= 714~d to 1136~d, inconsistent with the observed very stable late-time brightness.

AT2018cow has also been suggested to arise from a low-mass star disrupted by an intermediate-mass black hole (\citealt{Liu2018, Kuin2019}; \citetalias{P19}). In such a TDE, the stellar debris is usually assumed to be swallowed within a few orbits, giving rise to a $t^{-5/3}$ declining light curve after peak \citep{Rees1988}. We might expect some late-time brightness if there is significant time delay for some of the debris to be accreted. In this case, a constant accretion rate is needed to explain the stable F555W/F814W brightness at 714--1136~d; the H$\alpha$ detection also requires the disrupted star to be hydrogen-rich.


\section{Summary}

In this paper we report the discovery of a late-time source of the FBOT AT2018cow at 714~d and 1136~d after its explosion. It has a very stable brightness between the two epochs and a very blue SED consistent with the Rayleigh-Jeans tail of a blackbody spectrum with temperature log($T$/K)~$>$ 4.7 and luminosity log($L$/$L_\odot$)~$>$ 7.0. Significant H$\alpha$ emission is also detected. This late-time source is unlikely an unrelated object in chance alignment, or due to a light echo of AT2018cow. We discussed other possible scenarios as stellar source(s) (companion stars, star cluster, or the survived progenitor star) or due to AT2018cow's late emission (CSM interaction, magnetar, or delayed accretion in a TDE). All these scenarios have some difficulties in explaining this late-time source and require some fine tuning or special mechanisms to match the observations. It cannot be ruled out, however, that the late-time brightness is contributed by multiple components and/or processes. We believe this late-time source may contain important information for AT2018cow, but long-term monitoring and multi-wavelength observations are required to resolve its origin.


\section*{Acknowledgements}

We thank the anonymous referee for the very helpful comments on our paper. Research of N-CS and JRM is funded by the Science and Technology Facilities Council through grant ST/V000853/1. L-DL is supported by the
National Key R\&D Program of China (2021YFA0718500). This paper is based on observations made with the NASA/ESA Hubble Space Telescope and has used the light curves published by \citetalias{P19}.

\section*{Data availability}

Data used in this work are all publicly available from the Mikulski Archive for Space Telescope (\url{https://archive.stsci.edu}) or from the paper by \citetalias{P19}.

\bsp	
\label{lastpage}
\end{document}